\documentclass[runningheads]{llncs}
\usepackage{graphicx}
\usepackage{booktabs}
\usepackage{amsfonts}
\usepackage{amsmath}
\usepackage{tikz}
\usetikzlibrary{shapes,snakes}
\begin{document}
\title{Greedy Algorithms for Decision Trees with Hypotheses}
\titlerunning{Greedy Algorithms for Decision Trees with Hypotheses}
%
\author{Mohammad Azad\inst{1}\orcidID{0000-0001-9851-1420} \and
Igor Chikalov\inst{2}\orcidID{0000-0002-1010-6605} \and
Shahid Hussain\inst{3}\orcidID{0000-0002-1698-2809} \and
Mikhail Moshkov\inst{4}\orcidID{0000-0003-0085-9483} \and
Beata Zielosko\inst{5}\orcidID{0000-0003-3788-1094}}
\authorrunning{M. Azad et al.}
%
\institute{Department of Computer Science, College of Computer and Information Sciences, Jouf University, Sakaka~72441, Saudi Arabia\\
\email{mmazad@ju.edu.sa}
\and
Intel Corporation, 5000 W Chandler Blvd, Chandler, AZ 85226, USA\\
\email{igor.chikalov@gmail.com}
\and
Department of Computer Science, School of Mathematics and Computer Science, Institute of Business Administration, University Road, Karachi 75270, Pakistan\\
\email{shahidhussain@iba.edu.pk}
\and
Computer, Electrical and Mathematical Sciences \& Engineering Division, King Abdullah University of Science and Technology, Thuwal 23955-6900, Saudi Arabia\\
\email{mikhail.moshkov@kaust.edu.sa}
\and
Institute of Computer Science, Faculty of Science and Technology, University of Silesia in Katowice, B\c{e}dzi\'{n}ska 39, 41-200 Sosnowiec, Poland\\
\email{beata.zielosko@us.edu.pl}
}
\maketitle              
\begin{abstract}
We investigate at decision trees that incorporate both traditional queries based on one attribute and queries based on hypotheses about the values of all attributes.  Such decision trees
are similar to ones studied in exact learning, where membership and
equivalence queries are allowed. We present greedy algorithms based
on diverse uncertainty measures for construction of above decision
trees and discuss results of computer experiments on various data
sets from  the UCI ML Repository and randomly generated Boolean functions. We also study the
length and coverage of decision rules derived from the decision
trees constructed by greedy algorithms.

\end{abstract}

\section{Introduction}

\label{S1}

Decision trees are well known as classifiers, as a tool for knowledge
representation, and as algorithms \cite{CART,book13,Moshkov05,Rokach}. Conventional
decision trees are studied, in particular, in rough set theory initiated by
Pawlak \cite{Pawlak82,Pawlak91,Pawlak07} and in test theory initiated by
Chegis and Yablonskii \cite{Chegis58}. These trees use simple queries based
on one attribute each. In contrast to these theories, exact learning
initiated by Angluin \cite{Angluin88,Angluin04} studied not only membership
queries that correspond to attributes from rough set theory and test theory
but also the so-called equivalence queries.

In \cite{ent,Azad21a,Azad21b,ele,ent1}, we added the notion of a hypothesis to
the model that has been considered in rough set theory as well in test
theory. This model allows us to use an analog of equivalence queries and to
consider different types of decision trees based on various combinations of
attributes and hypotheses.

Experimental results discussed in \cite{ele} show that the optimal decision
trees with hypotheses can have less complexity than the conventional
decision trees and can be used as a tool for knowledge representation.
However, dynamic programming algorithms for the optimization of decision
trees considered in \cite{ele} are too complicated to be used in practice.
Therefore in \cite{ent} we proposed as entropy-based greedy algorithm for
the construction of different types of decision trees.

The present paper is
a generalization of \cite{ent}. For an arbitrary uncertainty measure, we
propose a greedy algorithm that, for given decision table and type of
decision trees, constructs a decision tree of the considered type for this
table.

The first goal of the present paper is to understand which uncertainty
measures and types of decision trees should be chosen if we would like to
minimize the depth or the number of realizable nodes in the constructed decision trees. To this end, we
compare parameters of the decision trees of different types constructed for
10 decision tables from the UCI ML Repository \cite{UCI} using five
uncertainty measures. We do the same for $100$ randomly generated Boolean
functions with~$n$ variables, where $n=3,\ldots ,6$.

We also study the length and coverage of decision rules derived from the
decision trees constructed by greedy algorithms. Previously in \cite{ent1},
we studied decision rules derived from optimal decision trees constructed
by dynamic programming algorithms. The second goal of the
paper is to understand which uncertainty measures and types of decision
trees should be chosen if we would like to minimize the length or to
maximize the coverage of the derived decision rules.

The main contributions of the paper are (i) the design of the greedy
algorithms that can work with arbitrary uncertainty measures and different
types of decision trees and (ii) the understanding (based on the experimental results) which uncertainty
measures and which types of decision trees should be chosen if we would like
to optimize the decision trees constructed by greedy algorithms or the decision rules derived from these trees.

The obtained experimental results  for Boolean functions do not depend on the used
uncertainty measures. We found the explanation of  this interesting fact.

The rest of the paper is organized as follows. In Section \ref{S2}, we
consider main notions and in Section \ref{S3}---greedy algorithms for the
decision tree construction. Sections \ref{S4}--\ref{S6} contain results of
computer experiments and their analysis, and Section \ref{S7}---short
conclusions.

\section{Main Notions}

\label{S2}

Detailed definitions related to the decision tables, decision trees, and decision rules can
be found in \cite{ent,ent1}. In this section, we restrict ourselves to the
necessary short comments. We also add some new definitions compared to the
papers \cite{ent,ent1}.

Let $T$ be a decision table with $n$ conditional attributes $f_{1},\ldots
,f_{n}$ having values from the set $\omega =\{0,1,2,\ldots \}$ in which rows
are pairwise different and each row is labeled with a decision from $\omega $%
. This table is called degenerate if it is empty or all rows of $T$ are
labeled with the same decision. We denote $F(T)=\{f_{1},\ldots ,f_{n}\}$ and
$D(T)$ the set of decisions attached to rows of $T$. For $f_{i}\in F(T)$, we
denote by $E(T,f_{i})$ the set of values of the attribute $f_{i}$ in the
table $T$.

A subtable of the table $T$ is a decision table obtained from $T$ by removal
of some rows. Let $S$ be a system of equations of the kind $f_{i}=\delta $
where $\delta \in E(T,f_{i})$. By $TS$ we denote a subtable of the table $T$
containing only rows satisfying all equations from $S$.

We denote by $N(T)$ the number of rows in $T$ and, for any $t\in D(T)$, we
denote by $N_{t}(T)$ the number of rows of $T$ labeled with the decision $t$%
. By $mcd(T)$ we denote a most common decision for $T$. If $T$ is empty,
then $mcd(T)=0$.

We denote by $\mathcal{T}$ the set of all decision tables. An uncertainty
measure is a function $U:\mathcal{T}\rightarrow \mathbb{R}$ such that $%
U(T)\geq 0$ for any $T\in \mathcal{T}$, and $U(T)=0$ if and only if $T$ is a
degenerate table. One can show (see book \cite{book19}) that the following
functions (we assume that, for any empty table, the value of each of the
considered functions is equal to $0$) are uncertainty measures:

\begin{itemize}
\item Misclassification error $me(T)=N(T)-N_{mcd(T)}(T)$.

\item Relative misclassification error $rme(T)=(N(T)-N_{mcd(T)}(T))/N(T)$.

\item Entropy $ent(T)=-\sum_{t\in D(T)}(N_{t}(T)/N(T))\log
_{2}(N_{t}(T)/N(T))$.

\item Gini index $gini(T)=1-\sum_{t\in D(T)}(N_{t}(T)/N(T))^{2}$.

\item Function $R$, where $R(T)$ is the number of unordered pairs of rows of
$T$ labeled with different decisions (note that $R(T)=N(T)^{2}gini(T)/2$).
\end{itemize}

For a given row of $T$, we should recognize the decision attached to this
row. To this end, we can use decision trees based on two types of queries.
We can ask about the value of an attribute $f_{i}$ on the given row. This
query has the set of answers $A(f_{i})=\{\{f_{i}=\delta \}:\delta \in
E(T,f_{i})\}$. We can formulate a hypothesis over $T$ in the form of $%
H=\{f_{1}=\delta _{1},\ldots ,f_{n}=\delta _{n}\}$, where $\delta _{1}\in
E(T,f_{1}),\ldots ,\delta _{n}\in E(T,f_{n})$, and ask about this
hypothesis. This query has the set of answers $A(H)=\{H,\{f_{1}=\sigma
_{1}\},...,\{f_{n}=\sigma _{n}\}:\sigma _{1}\in E(T,f_{1})\setminus \{\delta
_{1}\},...,\sigma _{n}\in E(T,f_{n})\setminus \{\delta _{n}\}\}$. The answer
$H$ means that the hypothesis is true. Other answers are counterexamples.
The hypothesis $H$ is called proper for $T$ if $(\delta _{1},\ldots ,\delta
_{n})$ is a row of the table $T$.

In this paper, we consider the following five types of decision trees:

\begin{enumerate}
\item Decision trees that use only attributes.

\item Decision trees that use only hypotheses.

\item Decision trees that use both attributes and hypotheses.

\item Decision trees that use only proper hypotheses.

\item Decision trees that use both attributes and proper hypotheses.
\end{enumerate}

We consider the depth $h(\Gamma)$ of a decision tree $\Gamma$ as its time
complexity, which is equal to the maximum number of queries in a path from
the root to a terminal node of the tree. As the space complexity of a
decision tree $\Gamma$, we consider the number of its realizable relative to
$T$ nodes $L(T,\Gamma)$. A node is called realizable relative to $T$ if, for
a row of $T$ and some choice of counterexamples, the computation in the tree
will pass through this node.

A complete path $\xi $ in $\Gamma $ is an arbitrary directed path from the
root to a terminal node. Denote $T(\xi )=TS(\xi )$, where $S(\xi )$ is the
union of systems of equations attached to edges of the path $\xi $.

Let $\Gamma $ be a decision tree for $T$, $\xi $ be a complete path in $%
\Gamma $ such that $T(\xi )$ is a nonempty table, and the terminal node of
the path $\xi $ be labeled with the decision $d$. We now define a system of
equations $S^{\prime }(\xi )$. If there are no working nodes in $\xi $, then
$S^{\prime }(\xi )$ is the empty system. Let us assume now that $\xi $
contains at least one working node. We now transform systems of equations
attached to edges leaving working nodes of $\xi $. If an edge is labeled
with an equation system containing exactly one equation, then we will not
change this system. Let an edge $e$ leaving a working node $v$ be labeled
with an equation system containing more than one equation. Then $v$ is
labeled with a hypothesis $H$ and $e$ is labeled with the equation system $H$%
. Note that if such a node exists, then it is the last working node in the
complete path $\xi $. In this case, we remove from the equation system $H$
attached to $e$ all equations, which follow from the union of equation
systems attached to edges of the path from the root to the node $v$. Then $%
S^{\prime }(\xi )$ is the union of new equation systems attached to the
edges of the path $\xi $. Note that the removed equations are redundant: $%
T(\xi )=TS^{\prime }(\xi )$.

We correspond to the complete path $\xi $ the decision rule
\[
\bigwedge_{f_{i}=\delta \in S^{\prime }(\xi )}(f_{i}=\delta )\rightarrow d.
\]%
We denote this rule by $rule(\xi )$. The number of equations in the equation
system $S^{\prime }(\xi )$ is called the length of the rule $rule(\xi )$ and
is denoted $l(rule(\xi ))$. The number of rows in the subtable $T(\xi )$ is
called the coverage of the rule $rule(\xi )$ and is denoted $c(rule(\xi ))$.

Denote $\Xi (T,\Gamma )$ the set of complete paths $\xi $ in $\Gamma $ such
that the table $T(\xi )$ is nonempty and $Rows(T)$ the set of rows of the
decision table $T$. For a row $r\in Rows(T)$, we denote by $l(r,T,\Gamma )$
the minimum length of a rule $rule(\xi )$ such that $\xi \in \Xi (T,\Gamma )$
and $r$ is a row of the subtable $T(\xi )$, and we denote by $c(r,T,\Gamma )$
the maximum coverage of a rule $rule(\xi )$ such that $\xi \in \Xi (T,\Gamma
)$ and $r$ is a row of the subtable $T(\xi )$.

We will use the following notation:

\begin{eqnarray*}
l(T,\Gamma ) &=&\frac{\sum_{r\in Rows(T)}l(r,T,\Gamma )}{|Rows(T)|}, \\
c(T,\Gamma ) &=&\frac{\sum_{r\in Rows(T)}c(r,T,\Gamma )}{|Rows(T)|}.
\end{eqnarray*}

\section{Greedy Algorithms}

\label{S3}

Let $U$ be an uncertainty measure, $T$ be a nondegenerate decision table
with $n$ conditional attributes $f_{1},\ldots ,f_{n}$, and $\Theta $ be a
nondegenerate subtable of the table $T$. We now define the impurity of a
query for the table $\Theta $ and uncertainty measure $U$. The impurity of
the query based on an attribute $f_{i}\in F(T)$ (impurity of query $f_{i}$)
is equal to $I_{U}(f_{i},\Theta )=\max \{U(\Theta S):S\in A(f_{i})\}$. The
impurity of the query based on a hypothesis $H$ (impurity of query $H$) is
equal to $I_{U}(H,\Theta )=\max \{U(\Theta S):S\in A(H)\}$.

An attribute $f_{i}$ is called admissible for $\Theta $ if it is not
constant in $\Theta $. A hypothesis $\{f_{1}=\delta _{1},\ldots
,f_{n}=\delta _{n}\}$ over $T$ is called admissible for $\Theta $ if it
satisfies the following condition. For $i=1,\ldots ,n$, if $f_{i}$ is
constant in $\Theta $, then $\delta _{i}$ is the only value of $f_{i}$ in $%
\Theta $.

We can find by simple search among all attributes an admissible for $\Theta $
attribute $f_{i}$ with the minimum impurity $I_{U}(f_{i},\Theta )$. We can
also find by simple search among all proper hypotheses an admissible for $%
\Theta $ proper hypothesis $H$ with the minimum impurity $I_{U}(H,\Theta )$.
It is not necessary to consider all hypotheses to find an admissible for $%
\Theta $ hypothesis with the minimum impurity. For $i=1,\ldots ,n$, we
denote by $\delta _{i}$ a number from $E(T,f_{i})$ such that $U(\Theta
\{f_{i}=\delta _{i}\})=\max \{U(\Theta \{f_{i}=\sigma \}):\sigma \in
E(T,f_{i})\}$. Then the hypothesis $H=\{f_{1}=\delta _{1},\ldots
,f_{n}=\delta _{n}\}$ is admissible for $\Theta $ and has the minimum
impurity $I_{U}(H,\Theta )$ among all admissible for $\Theta $ hypotheses.

We now describe a greedy algorithm $\mathcal{A}_{U}$ based on the
uncertainty measure $U$ that, for a given nonempty decision table $T$ and $%
k\in \{1,\ldots ,5\}$, constructs a decision tree of type $k$ for the table $%
T$. This algorithm is a generalization of the algorithm considered in \cite{ent}.
\medskip

\noindent {\bf Algorithm} $\mathcal{A}_U$.

\noindent\emph{{Input} }: A nonempty decision table $T$ and a number $k\in
\{1,\ldots ,5\}$.

\noindent \emph{Output}: A decision tree of  type $k$ for the table $T$.

\begin{enumerate}
\item Construct a tree $G$ consisting of a single node labeled with $T$.

\item If no node of the tree $G$ is labeled with a table, then the algorithm
ends and returns the tree $G$.

\item Choose a node $v$ in $G$, which is labeled with a subtable $\Theta $
of the table $T$.

\item If $\Theta $ is degenerate, then instead of $\Theta $, we label the
node $v$ with $0$ if $\Theta $ is empty and with the decision attached to
each row of $\Theta $ if $\Theta $ is nonempty.

\item If $\Theta $ is nondegenerate, then depending on $k$ we choose an
admissible for $\Theta $ query $X$ (either attribute or hypothesis) in the
following way:

\begin{enumerate}
\item If $k=1$, then we find an admissible for $\Theta $ attribute $X$ $\in
F(T)$ with the minimum impurity $I_{U}(X,\Theta )$.

\item If $k=2$, then we find an admissible for $\Theta $ hypothesis $X$ over
$T$ with the minimum impurity $I_{U}(X,\Theta )$.

\item If $k=3$, then we find an admissible for $\Theta $ attribute $Y$ $\in
F(T)$ with the minimum impurity $I_{U}(Y,\Theta )$ and an admissible for $%
\Theta $ hypothesis $Z$ over $T$ with the minimum impurity $I_{U}(Z,\Theta )$%
. Between $Y$ and $Z$, we choose a query $X$ with the minimum impurity $%
I_{U}(X,\Theta )$.

\item If $k=4$, then we find an admissible for $\Theta $ proper hypothesis $X
$ over $T$ with the minimum impurity $I_{U}(X,\Theta )$.

\item If $k=5$, then we find an admissible for $\Theta $ attribute $Y$ $\in
F(T)$ with the minimum impurity $I_{U}(Y,\Theta )$ and an admissible for $%
\Theta $ proper hypothesis $Z$ over $T$ with the minimum impurity $%
I_{U}(Z,\Theta )$. Between $Y$ and $Z$, we choose a query $X$ with the
minimum impurity $I_{U}(X,\Theta )$.
\end{enumerate}

\item Instead of $\Theta $, we label the node $v$ with the query $X$. For
each answer $S\in A(X)$, we add to the tree $G$ a node $v(S)$ and an edge $%
e(S)$ connecting $v$ and $v(S)$. We label the node $v(S)$ with the subtable $%
\Theta S$ and label the edge $e(S)$ with the answer $S$. Proceed to step 2.
\end{enumerate}
\medskip

For a given nonempty decision table $T$ and number $k\in \{1,\ldots ,5\}$,
the algorithm $\mathcal{A}_{U}$ constructs a decision tree $\Gamma $ of
type $k$ for the table $T$. We will use the following notation:

\begin{eqnarray*}
h_{U}^{(k)}(T) &=&h(\Gamma ), \\
L_{U}^{(k)}(T) &=&L(T,\Gamma ), \\
l_{U}^{(k)}(T) &=&l(T,\Gamma ), \\
c_{U}^{(k)}(T) &=&c(T,\Gamma ).
\end{eqnarray*}

\section{Results of Experiments with Decision Tables from \cite{UCI}}

\label{S4}

We now consider results of experiments with decision tables described in Table \ref{tab0}.

\begin{table}[h!]
\caption{Decision tables from \protect\cite{UCI} used in experiments}
\label{tab0}
\begin{tabular}{ccc}
\toprule Decision & Number of & Number of  \\
table & rows & attributes  \\
\midrule \textsc{balance-scale} & 625 & 5  \\
\textsc{breast-cancer} & 266 & 10  \\
\textsc{cars} & 1728 & 7  \\
\textsc{hayes-roth-data} & 69 & 5  \\
\textsc{lymphography} &	148 &	18 \\
\textsc{nursery} & 12960 & 9  \\
\textsc{soybean-small} & 47 & 36 \\
\textsc{spect-test} &	169 &	22 \\
\textsc{tic-tac-toe} & 958 & 10  \\
\textsc{zoo-data} & 59 & 17  \\
\bottomrule &  &
\end{tabular}%
\end{table}

Using the algorithm $\mathcal{A}_{U}$ with five uncertainty measures, we construct for these tables  decision trees of different types, evaluate complexity of these trees and study decision rules derived from them.

\subsection{Results for Misclassification Error $me$}

Results for decision trees can be found in Tables \ref{tab1.1} and \ref{tab1.2}. In
Table \ref{tab1.1}, we consider parameters $h_{me}^{(1)}(T),\ldots ,h_{me}^{(5)}(T)$
(minimum values  are in bold).

\begin{table}[h!]
\caption{Results for  $me$ and $h$}
\label{tab1.1}
\begin{tabular}{cccccc}
\toprule  Decision  &   $h_{me}^{(1)}(T)$   &   $h_{me}^{(2)}(T)$ &   $%
h_{me}^{(3)}(T)$ &   $h_{me}^{(4)}(T)$ &   $h_{me}^{(5)}(T)$ \\
 table   $T$  &  &  &  &  &  \\
\midrule
\sc balance-scale   & \bf 4   & \bf 4   & \bf 4   & \bf 4 & \bf 4   \\
\sc breast-cancer   & 7   & 6   & \bf 5   & 6 & \bf 5   \\
\sc cars            & \bf 6   & \bf 6   & \bf 6   & \bf 6 & \bf 6   \\
\sc hayes-roth-data & \bf 4   & \bf 4   & \bf 4   & \bf 4 & \bf 4   \\
\sc lymphography    & 7   & \bf 5   & \bf 5   & 7 & \bf 5   \\
\sc nursery         & \bf 8   & \bf 8   & \bf 8   & \bf 8 & \bf 8   \\
\sc soybean-small   & \bf 2   & 5   & \bf 2   & 7 & \bf 2   \\
\sc spect-test      & 15  & \bf 4   & \bf 4   & \bf 4 & \bf 4   \\
\sc tic-tac-toe     & \bf 7   & \bf 7   & \bf 7   & 8 & \bf 7   \\
\sc zoo-data        & \bf 4   & \bf 4   & \bf 4   & 6 & \bf 4   \\ \midrule
Average         & 6.4 & 5.3 & 4.9 & 6 & 4.9\\
\bottomrule
\end{tabular}%
\end{table}

In Table \ref{tab1.2}, we consider parameters $%
L_{me}^{(1)}(T),\ldots ,L_{me}^{(5)}(T)$ (minimum values  are in bold).

\begin{table}[h!]
\caption{Results for $me$ and $L$}
\label{tab1.2}
\begin{tabular}{cccccc}
\toprule  Decision &   $L_{me}^{(1)}(T)$ &   $L_{me}^{(2)}(T)$ &   $L_{me}^{(3)}(T)$ &   $L_{me}^{(4)}(T)$
&   $L_{me}^{(5)}(T)$ \\
 table   $T$ &  &  &  &  &  \\
\midrule
\sc balance-scale   & \bf 556   & 5,234      & 3,694   & 5,234      & 3,694   \\
\sc breast-cancer   & \bf 238   & 21,922     & 285    & 33,099     & 254    \\
\sc cars            & \bf 1,462  & 66,593     & 1,590   & 66,593     & 1,590   \\
\sc hayes-roth-data & \bf 65    & 353       & 88     & 348       & 88     \\
\sc lymphography    & \bf 92    & 49,780     & 126    & 165,481    & 126    \\
\sc nursery         & \bf 4,623  & 13,487,465  & 5,473   & 13,487,465  & 5,473   \\
\sc soybean-small   & \bf 23    & 5,375      & \bf 23     & 44,601     & \bf 23     \\
\sc spect-test      & \bf 91    & 3,266      & 168    & 6,229      & 168    \\
\sc tic-tac-toe     & \bf 547   & 287,504    & 61,315  & 703,982    & 613    \\
\sc zoo-data        & \bf 27    & 1,482      & \bf 27     & 4,411      & \bf 27     \\ \midrule
Average         & 772.4 & 1,392,897.4 & 7,278.9 & 1,451,744.3 & 12,05.6\\
\bottomrule
\end{tabular}%
\end{table}

Results for decision rules can be found  in Tables \ref{tab1.5} and \ref{tab1.6}. In
Table \ref{tab1.5}, we consider parameters $l_{me}^{(1)}(T),\ldots ,l_{me}^{(5)}(T)$
(minimum values  are in bold).

\begin{table}[h!]
\caption{Results for $me$ and $l$}
\label{tab1.5}
\begin{tabular}{cccccc}
\toprule  Decision  &   $l_{me}^{(1)}(T)$   &   $l_{me}^{(2)}(T)$ &   $%
l_{me}^{(3)}(T)$ &   $l_{me}^{(4)}(T)$ &   $l_{me}^{(5)}(T)$ \\
 table   $T$  &  &  &  &  &  \\
\midrule
\sc balance-scale & 3.64 & \textbf{3.20} & 3.24 & \textbf{3.20} & 3.24\\
\sc breast-cancer & 3.61 &  \textbf{2.76} & 3.48 & 2.79 & 3.55\\
\sc cars & 5.05 &  \textbf{2.47} & 4.99 &  \textbf{2.47} & 4.99\\
\sc hayes-roth-data & 2.87 &  2.26 & 2.86 &  \textbf{2.25} & 2.86\\
\sc lymphography & 3.48 &  \textbf{1.99} & 3.53 & 2.14 & 3.53\\
\sc nursery & 4.82 &  \textbf{3.34} & 4.69 &  \textbf{3.34} & 4.69\\
\sc soybean-small & 1.89 &  \textbf{1.00} & 1.89 & 1.57 & 1.89\\
\sc spect-test & 5.44 & 2.27 & 4.81 &  \textbf{2.05} & 4.81\\
\sc tic-tac-toe & 5.48 & 3.35 & 3.36 &  \textbf{3.19} & 5.47\\
\sc zoo-data & 2.53 &  \textbf{1.56} & 2.53 & 1.85 & 2.53\\ \midrule
Average & 3.88 & 2.42 & 3.54 & 2.48 & 3.76\\
\bottomrule
\end{tabular}%
\end{table}

In Table \ref{tab1.6}, we consider parameters $%
c_{me}^{(1)}(T),\ldots ,c_{me}^{(5)}(T)$ (maximum values  are in bold).

\begin{table}[h!]
\caption{Results for $me$ and $c$}
\label{tab1.6}
\begin{tabular}{cccccc}
\toprule  Decision &   $c_{me}^{(1)}(T)$ &   $c_{me}^{(2)}(T)$ &   $c_{me}^{(3)}(T)$ &   $c_{me}^{(4)}(T)$
&   $c_{me}^{(5)}(T)$ \\
 table   $T$ &  &  &  &  &  \\
\midrule
\sc balance-scale & 2.44 & \textbf{4.21} & 4.05 & \textbf{4.21} & 4.05\\
\sc breast-cancer & 4.72 & 8.38 & 5.16 & \textbf{8.84} & 4.97\\
\sc cars & 5.74 & \textbf{332.63} & 5.97 & \textbf{332.63} & 5.97\\
\sc hayes-roth-data & 3.58 & \textbf{6.19} & 3.62 & \textbf{6.19} & 3.62\\
\sc lymphography & 4.14 & 18.19 & 4.51 & \textbf{18.70} & 4.51\\
\sc nursery & 300.54 & \textbf{1,523.16} & 304.06 & \textbf{1,523.16} & 304.06\\
\sc soybean-small & 3.47 & \textbf{12.32} & 3.47 & 9.32 & 3.47\\
\sc spect-test & 8.75 & \textbf{57.98} & 18.69 & 57.87 & 18.69\\
\sc tic-tac-toe & 6.31 & 49.52 & 50.26 & \textbf{56.26} & 6.70\\
\sc zoo-data & 6.83 & 10.59 & 6.83 & \textbf{10.80} & 6.83\\
\midrule
Average & 34.65 & 202.32 & 40.66 & 202.80 & 36.29\\
\bottomrule
\end{tabular}%
\end{table}

\subsection{Results for Relative Misclassification Error $rme$}

Results for decision trees can be found  in Tables \ref{tab2.1} and \ref{tab2.2}. In
Table \ref{tab2.1},  we consider parameters $h_{rme}^{(1)}(T),\ldots ,h_{rme}^{(5)}(T)$
(minimum values  are in bold).

\begin{table}[h!]
\caption{Results for $rme$ and $h$}
\label{tab2.1}
\begin{tabular}{cccccc}
\toprule  Decision  &   $h_{rme}^{(1)}(T)$   &   $h_{rme}^{(2)}(T)$ &   $%
h_{rme}^{(3)}(T)$ &   $h_{rme}^{(4)}(T)$ &   $h_{rme}^{(5)}(T)$ \\
 table   $T$  &  &  &  &  &  \\
\midrule
\sc balance-scale   & \bf 4   & \bf 4   & \bf 4   & \bf 4   & \bf 4   \\
\sc breast-cancer   & 9   & 9   & 9   & \bf 8   & 9   \\
\sc cars            & \bf 6   &\bf  6   & \bf 6   & \bf 6   & \bf 6   \\
\sc hayes-roth-data & \bf 4   & \bf 4   & \bf 4   & \bf 4   & \bf 4   \\
\sc lymphography    & \bf 10  & 11  & \bf 10  & 12  & \bf 10  \\
\sc nursery         & \bf 8   & \bf 8   & \bf 8   & \bf 8   & \bf 8   \\
\sc soybean-small   & \bf 2   & 6   & \bf 2   & 8   & \bf 2   \\
\sc spect-test      & 20  & \bf 5   & \bf 5   & 14  & 11  \\
\sc tic-tac-toe     & \bf 7   & 8   & \bf 7   & 8   & \bf 7   \\
\sc zoo-data        & 8   & 7   & \bf 6   & 10  & \bf 6   \\ \midrule
Average         & 7.8 & 6.8 & 6.1 & 8.2 & 6.7\\
\bottomrule
\end{tabular}%
\end{table}

In Table \ref{tab2.2}, we consider parameters $%
L_{rme}^{(1)}(T),\ldots ,L_{rme}^{(5)}(T)$ (minimum values  are in bold).

\begin{table}[h!]
\caption{Results for $rme$ and $L$}
\label{tab2.2}
\begin{tabular}{cccccc}
\toprule  Decision &   $L_{rme}^{(1)}(T)$ &   $L_{rme}^{(2)}(T)$ &   $L_{rme}^{(3)}(T)$ &   $L_{rme}^{(4)}(T)$
&   $L_{rme}^{(5)}(T)$ \\
 table   $T$ &  &  &  &  &  \\
\midrule
\sc balance-scale   & \bf 556   & 5,234     & 3,694    & 5,234      & 3,694   \\
\sc breast-cancer   & \bf 255   & 446,170   & 304     & 103,642    & 266    \\
\sc cars            & \bf 1,592  & 66,593    & 1,831    & 66,593     & 1,831   \\
\sc hayes-roth-data & 73    & 428      & \bf 72      & 378       & \bf 72     \\
\sc lymphography    & \bf 116   & 2,475,650  & 143     & 1,952,599   & 143    \\
\sc nursery         & \bf 4,493  & 13,487,465 & 13,667   & 13,487,465  & 13,667  \\
\sc soybean-small   & \bf 7     & 11,403    & \bf 7       & 113,855    & \bf 7      \\
\sc spect-test      & \bf 123   & 6,983     & 5,495    & 398,926    & 1,116   \\
\sc tic-tac-toe     & \bf 648   & 864,578   & 200,847  & 946,858    & 940    \\
\sc zoo-data        & \bf 35    & 6,536     & 37      & 30,889     & 37     \\ \midrule
Average         & 789.8 & 1,737,104  & 22,609.7 & 1,710,643.9 & 2,177.3 \\
\bottomrule
\end{tabular}%
\end{table}

Results for decision rules can be found  in Tables \ref{tab2.5} and \ref{tab2.6}. In
Table \ref{tab2.5}, we consider parameters $l_{rme}^{(1)}(T),\ldots ,l_{rme}^{(5)}(T)$
(minimum values  are in bold).

\begin{table}[h!]
\caption{Results for $rme$ and $l$}
\label{tab2.5}
\begin{tabular}{cccccc}
\toprule  Decision  &   $l_{rme}^{(1)}(T)$   &   $l_{rme}^{(2)}(T)$ &   $%
l_{rme}^{(3)}(T)$ &   $l_{rme}^{(4)}(T)$ &   $l_{rme}^{(5)}(T)$ \\
 table   $T$  &  &  &  &  &  \\
\midrule
\sc balance-scale & 3.64 & \textbf{3.20} & 3.24 & \textbf{3.20} & 3.24\\
\sc breast-cancer & 6.11 & \textbf{2.68} & 6.01 & 2.73 & 6.06\\
\sc cars & 5.30 & \textbf{2.47} & 5.18 & \textbf{2.47} & 5.18\\
\sc hayes-roth-data & 3.22 & \textbf{2.20} & 3.20 & 2.22 & 3.20\\
\sc lymphography & 6.97 & \textbf{2.01} & 6.80 & 2.12 & 6.80\\
\sc nursery & 4.72 & \textbf{3.34} & 4.33 & \textbf{3.34} & 4.33\\
\sc soybean-small & 1.34 & \textbf{1.00} & 1.34 & 1.53 & 1.34\\
\sc spect-test & 8.37 & 2.28 & 2.18 & \textbf{1.79} & 2.21\\
\sc tic-tac-toe & 5.77 & 3.52 & 3.70 & \textbf{3.19} & 5.63\\
\sc zoo-data & 4.90 & \textbf{1.66} & 4.61 & 1.83 & 4.61\\\midrule
Average & 5.03 & 2.43 & 4.06 & 2.44 & 4.26\\
\bottomrule
\end{tabular}%
\end{table}

In Table \ref{tab2.6}, we consider parameters $%
c_{rme}^{(1)}(T),\ldots ,c_{rme}^{(5)}(T)$ (maximum values  are in bold).

\begin{table}[h!]
\caption{Results for $rme$ and $c$}
\label{tab2.6}
\begin{tabular}{cccccc}
\toprule  Decision &   $c_{rme}^{(1)}(T)$ &   $c_{rme}^{(2)}(T)$ &   $c_{rme}^{(3)}(T)$ &   $c_{rme}^{(4)}(T)$
&   $c_{rme}^{(5)}(T)$ \\
 table   $T$ &  &  &  &  &  \\
\midrule
\sc balance-scale & 2.44 & \textbf{4.21} & 4.05 & \textbf{4.21} & 4.05\\
\sc breast-cancer & 2.18 & 9.12 & 2.68 & \textbf{9.46} & 2.37\\
\sc cars & 3.95 & \textbf{332.63} & 5.40 & \textbf{332.63} & 5.40\\
\sc hayes-roth-data & 1.81 & \textbf{6.23} & 1.81 & 6.22 & 1.81\\
\sc lymphography & 3.32 & 19.76 & 4.12 & \textbf{21.33} & 4.12\\
\sc nursery & 1,451.06 & \textbf{1,523.16} & 1,460.80 & \textbf{1,523.16} & 1,460.80\\
\sc soybean-small & 11.51 & \textbf{12.32} & 11.51 & 11.11 & 11.51\\
\sc spect-test & 5.54 & 57.98 & 57.98 & \textbf{58.01} & 54.76\\
\sc tic-tac-toe & 5.31 & 50.48 & 34.99 & \textbf{56.25} & 6.27\\
\sc zoo-data & 7.10 & 10.69 & 7.71 & \textbf{10.97} & 7.71\\
\midrule
Average & 149.42 & 202.66 & 159.10 & 203.33 & 155.88\\
\bottomrule
\end{tabular}%
\end{table}

\subsection{Results for Entropy $ent$}

For the completeness, we repeat experiments considered in \cite{ent} and related to entropy and decision trees.

Results for decision trees can be found  in Tables \ref{tab3.1} and \ref{tab3.2}. In
Table \ref{tab3.1}, we consider parameters $h_{ent}^{(1)}(T),\ldots ,h_{ent}^{(5)}(T)$
(minimum values  are in bold).

\begin{table}[h!]
\caption{Results for $ent$ and $h$}
\label{tab3.1}
\begin{tabular}{cccccc}
\toprule  Decision  &   $h_{ent}^{(1)}(T)$   &   $h_{ent}^{(2)}(T)$ &   $%
h_{ent}^{(3)}(T)$ &   $h_{ent}^{(4)}(T)$ &   $h_{ent}^{(5)}(T)$ \\
 table   $T$  &  &  &  &  &  \\
\midrule
\sc balance-scale   & \bf 4   & \bf 4   & \bf 4   & \bf 4   & \bf 4   \\
\sc breast-cancer   & 9   & 9   & 9   & \bf 8   & 9   \\
\sc cars            & \bf 6   & \bf 6   & \bf 6   & \bf 6   & \bf 6   \\
\sc hayes-roth-data & \bf 4   & \bf 4   & \bf 4   & \bf 4   & \bf 4   \\
\sc lymphography    & 11  & 11  & \bf 9   & 13  & 10  \\
\sc nursery         & \bf 8   & \bf 8   & \bf 8   & \bf 8   & \bf 8   \\
\sc soybean-small   & \bf 2   & 6   & \bf 2   & 8   & \bf 2   \\
\sc spect-test      & 20  & \bf 5   & \bf 5   & 14  & 11  \\
\sc tic-tac-toe     & \bf 7   & 8   & \bf 7   & 8   & \bf 7   \\
\sc zoo-data        & 8   & 6   & \bf 5   & 8   & \bf 5   \\ \midrule
Average         & 7.9 & 6.7 & 5.9 & 8.1 & 6.6 \\
\bottomrule
\end{tabular}%
\end{table}

In Table \ref{tab3.2}, we consider parameters  $%
L_{ent}^{(1)}(T),\ldots ,L_{ent}^{(5)}(T)$ (minimum values  are in bold).

\begin{table}[h!]
\caption{Results for $ent$ and $L$}
\label{tab3.2}
\begin{tabular}{cccccc}
\toprule  Decision &   $L_{ent}^{(1)}(T)$ &   $L_{ent}^{(2)}(T)$ &   $L_{ent}^{(3)}(T)$ &   $L_{ent}^{(4)}(T)$
&   $L_{ent}^{(5)}(T)$ \\
 table   $T$ &  &  &  &  &  \\
\midrule
\sc balance-scale   & \bf 556   & 5,234      & 4,102   & 5,234      & 4,102   \\
\sc breast-cancer   & \bf 255   & 446,170    & 304    & 103,642    & 266    \\
\sc cars            & \bf 1,136  & 65,624     & 3,944   & 65,624     & 3,944   \\
\sc hayes-roth-data & 73    & 421       & \bf 72     & 367       & \bf 72     \\
\sc lymphography    & \bf 123   & 6,653,366   & 162    & 8,515,841   & 153    \\
\sc nursery         & \bf 4,460  & 12,790,306  & 14,422  & 12,790,306  & 14,422  \\
\sc soybean-small   & \bf 7     & 10,029     & \bf 7      & 157,640    & \bf 7      \\
\sc spect-test      & \bf 123   & 6,983      & 5,495   & 398,926    & 1,116   \\
\sc tic-tac-toe     & \bf 648   & 864,578    & 200,847 & 946,858    & 940    \\
\sc zoo-data        & \bf 33    & 2,134      & 35     & 13,310     & 35     \\ \midrule
Average         & 741.4 & 2,084,484.5 & 22,939  & 2,299,774.8 & 2,505.7\\
\bottomrule
\end{tabular}%
\end{table}

Results for decision rules can be found  in Tables \ref{tab3.5} and \ref{tab3.6}. In
Table \ref{tab3.5}, we consider parameters $l_{ent}^{(1)}(T),\ldots ,l_{ent}^{(5)}(T)$
(minimum values  are in bold).

\begin{table}[h!]
\caption{Results for $ent$ and $l$}
\label{tab3.5}
\begin{tabular}{cccccc}
\toprule  Decision  &   $l_{ent}^{(1)}(T)$   &   $l_{ent}^{(2)}(T)$ &   $%
l_{ent}^{(3)}(T)$ &   $l_{ent}^{(4)}(T)$ &   $l_{ent}^{(5)}(T)$ \\
 table   $T$  &  &  &  &  &  \\
\midrule
\sc balance-scale & 3.64 & \textbf{3.20} & 3.24 & \textbf{3.20} & 3.24\\
\sc breast-cancer & 6.11 & \textbf{2.68} & 6.01 & 2.73 & 6.06\\
\sc cars & 4.68 & \textbf{2.45} & 4.17 & \textbf{2.45} & 4.17\\
\sc hayes-roth-data & 3.22 & \textbf{2.17} & 3.20 & 2.22 & 3.20\\
\sc lymphography & 6.61 & \textbf{1.99} & 6.32 & 2.11 & 6.37\\
\sc nursery & 4.72 & \textbf{3.80} & 4.29 & \textbf{3.80} & 4.29\\
\sc soybean-small & 1.34 & \textbf{1.00} & 1.34 & 1.47 & 1.34\\
\sc spect-test & 8.37 & 2.28 & 2.18 & \textbf{1.79} & 2.21\\
\sc tic-tac-toe & 5.77 & 3.52 & 3.70 & \textbf{3.19} & 5.63\\
\sc zoo-data & 4.27 & \textbf{1.66} & 3.98 & 1.97 & 3.98\\
\midrule
Average & 4.87 & 2.48 & 3.84 & 2.49 & 4.05\\
\bottomrule
\end{tabular}%
\end{table}

In Table \ref{tab3.6}, we consider parameters $%
c_{ent}^{(1)}(T),\ldots ,c_{ent}^{(5)}(T)$ (maximum values  are in bold).

\begin{table}[h!]
\caption{Results for $ent$ and $c$}
\label{tab3.6}
\begin{tabular}{cccccc}
\toprule  Decision &   $c_{ent}^{(1)}(T)$ &   $c_{ent}^{(2)}(T)$ &   $c_{ent}^{(3)}(T)$ &   $c_{ent}^{(4)}(T)$
&   $c_{ent}^{(5)}(T)$ \\
 table   $T$ &  &  &  &  &  \\
\midrule
\sc balance-scale & 2.44 & \textbf{4.21} & 4.03 & \textbf{4.21} & 4.03\\
\sc breast-cancer & 2.18 & 9.12 & 2.68 & \textbf{9.46} & 2.37\\
\sc cars & 15.97 & \textbf{332.71} & 37.12 & \textbf{332.71} & 37.12\\
\sc hayes-roth-data & 1.81 & \textbf{6.38} & 1.81 & 6.26 & 1.81\\
\sc lymphography & 3.51 & \textbf{21.54} & 4.72 & 21.53 & 4.42\\
\sc nursery & 1,451.07 & \textbf{1,516.04} & 1,462.03 & \textbf{1,516.04} & 1,462.03\\
\sc soybean-small & 11.51 & \textbf{12.32} & 11.51 & 11.66 & 11.51\\
\sc spect-test & 5.54 & 57.98 & 57.98 & \textbf{58.01} & 54.76\\
\sc tic-tac-toe & 5.31 & 50.48 & 34.99 & \textbf{56.25} & 6.27\\
\sc zoo-data & 7.24 & \textbf{10.66} & 7.85 & 10.36 & 7.85\\
\midrule
Average & 150.66 & 202.14 & 162.47 & 202.65 & 159.22\\
\bottomrule
\end{tabular}%
\end{table}

\subsection{Results for Gini Index $gini$}

Results for decision trees can be found  in Tables \ref{tab4.1} and \ref{tab4.2}. In
Table \ref{tab4.1}, we consider parameters $h_{gini}^{(1)}(T),\ldots ,h_{gini}^{(5)}(T)$
(minimum values  are in bold).

\begin{table}[h!]
\caption{Results for $gini$ and $h$}
\label{tab4.1}
\begin{tabular}{cccccc}
\toprule  Decision  &   $h_{gini}^{(1)}(T)$   &   $h_{gini}^{(2)}(T)$ &   $%
h_{gini}^{(3)}(T)$ &   $h_{gini}^{(4)}(T)$ &   $h_{gini}^{(5)}(T)$ \\
 table   $T$  &  &  &  &  &  \\
\midrule
\sc balance-scale   & \bf 4   & \bf 4   & \bf 4   & \bf 4   & \bf 4   \\
\sc breast-cancer   & 9   & 9   & 9   & \bf 8   & 9   \\
\sc cars            & \bf 6   & \bf 6   & \bf 6   & \bf 6   & \bf 6   \\
\sc hayes-roth-data & \bf 4   & \bf 4   & \bf 4   & \bf 4   & \bf 4   \\
\sc lymphography    & \bf 10  & 11  & \bf 10  & 13  & \bf 10  \\
\sc nursery         & \bf 8   & \bf 8   & \bf 8   & \bf 8   & \bf 8   \\
\sc soybean-small   & \bf 2   & 6   & \bf 2   & 8   & \bf 2   \\
\sc spect-test      & 20  & \bf 5   & \bf 5   & 14  & 11  \\
\sc tic-tac-toe     & \bf 7   & 8   & \bf 7   & 8   & \bf 7   \\
\sc zoo-data        & 8   & \bf 7   & 8   & 9   & \bf 7   \\ \midrule
Average         & 7.8 & 6.8 & 6.3 & 8.2 & 6.8 \\
\bottomrule
\end{tabular}%
\end{table}

In Table \ref{tab4.2}, we consider parameters $%
L_{gini}^{(1)}(T),\ldots ,L_{gini}^{(5)}(T)$ (minimum values  are in bold).

\begin{table}[h!]
\caption{Results for $gini$ and $L$}
\label{tab4.2}
\begin{tabular}{cccccc}
\toprule  Decision &   $L_{gini}^{(1)}(T)$ &   $L_{gini}^{(2)}(T)$ &   $L_{gini}^{(3)}(T)$ &   $L_{gini}^{(4)}(T)$
&   $L_{gini}^{(5)}(T)$ \\
 table   $T$ &  &  &  &  &  \\
\midrule
\sc balance-scale   & \bf 556   & 5,234      & 4,006    & 5,234      & 4,006   \\
\sc breast-cancer   & \bf 255   & 446,170    & 304     & 103,642    & 266    \\
\sc cars            & \bf 1,511  & 65,579     & 3,643    & 65,579     & 3,643   \\
\sc hayes-roth-data & 73    & 415       & \bf 72      & 363       & \bf 72     \\
\sc lymphography    & \bf 115   & 2,735,180   & 140     & 2,232,312   & \bf 115    \\
\sc nursery         & 4,284  & 12,795,294  & 16,566   & 12,795,294  & 16,566  \\
\sc soybean-small   & \bf 7     & 10,029     & \bf 7       & 148,891    & \bf 7      \\
\sc spect-test      & \bf 123   & 6,983      & 5,495    & 398,926    & 1,116   \\
\sc tic-tac-toe     & \bf 648   & 864,578    & 200,847  & 946,858    & 940    \\
\sc zoo-data        & \bf 39    & 4,702      & 57      & 20,896     & 41     \\ \midrule
Average         & 761.1 & 1,693,416.4 & 23,113.7 & 1,671,799.5 & 2,677.2 \\
\bottomrule
\end{tabular}%
\end{table}

Results for decision rules can be found  in Tables \ref{tab4.5} and \ref{tab4.6}. In
Table \ref{tab4.5}, we consider parameters $l_{gini}^{(1)}(T),\ldots ,l_{gini}^{(5)}(T)$
(minimum values  are in bold).

\begin{table}[h!]
\caption{Results for $gini$ and $l$}
\label{tab4.5}
\begin{tabular}{cccccc}
\toprule  Decision  &   $l_{gini}^{(1)}(T)$   &   $l_{gini}^{(2)}(T)$ &   $%
l_{gini}^{(3)}(T)$ &   $l_{gini}^{(4)}(T)$ &   $l_{gini}^{(5)}(T)$ \\
 table   $T$  &  &  &  &  &  \\
\midrule
\sc balance-scale & 3.64 & \textbf{3.20} & 3.23 & \textbf{3.20} & 3.23\\
\sc breast-cancer  & 6.11 & \textbf{2.68} & 6.01 & 2.73 & 6.06\\
\sc cars  & 5.25 & \textbf{2.45} & 4.70 & \textbf{2.45} & 4.70\\
\sc hayes-roth-data  & 3.22 & \textbf{2.20} & 3.20 & 2.23 & 3.20\\
\sc lymphography  & 7.01 & \textbf{1.99} & 6.91 & 2.14 & 7.01\\
\sc nursery  & 4.63 & \textbf{3.76} & 4.12 & \textbf{3.76} & 4.12\\
\sc soybean-small  & 1.34 & \textbf{1.00} & 1.34 & 1.47 & 1.34\\
\sc spect-test  & 8.37 & 2.28 & 2.18 & \textbf{1.79} & 2.21\\
\sc tic-tac-toe & 5.77 & 3.52 & 3.70 & \textbf{3.19} & 5.63\\
\sc zoo-data & 5.10 & \textbf{1.66} & 4.71 & 2.05 & 4.81\\
\midrule
Average & 5.04 & 2.47 & 4.01 & 2.50 & 4.23\\
\bottomrule
\end{tabular}%
\end{table}

In Table \ref{tab4.6}, we consider parameters $%
c_{gini}^{(1)}(T),\ldots ,c_{gini}^{(5)}(T)$ (maximum values  are in bold).

\begin{table}[h!]
\caption{Results for $gini$ and $c$}
\label{tab4.6}
\begin{tabular}{cccccc}
\toprule  Decision &   $c_{gini}^{(1)}(T)$ &   $c_{gini}^{(2)}(T)$ &   $c_{gini}^{(3)}(T)$ &   $c_{gini}^{(4)}(T)$
&   $c_{gini}^{(5)}(T)$ \\
 table   $T$ &  &  &  &  &  \\
\midrule
\sc balance-scale & 2.44 & \textbf{4.21} & 4.08 & \textbf{4.21} & 4.08\\
\sc breast-cancer & 2.18 & 9.12 & 2.68 & \textbf{9.46} & 2.37\\
\sc cars & 4.41 & \textbf{332.70} & 26.27 & \textbf{332.70} & 26.27\\
\sc hayes-roth-data & 1.81 & \textbf{6.23} & 1.81 & 6.19 & 1.81\\
\sc lymphography & 3.28 & 20.85 & 3.49 & \textbf{21.45} & 3.28\\
\sc nursery & 1,454.14 & \textbf{1,516.81} & 1,467.10 &\textbf{1,516.81} & 1,467.10\\
\sc soybean-small & 11.51 & \textbf{12.32} & 11.51 & 11.66 & 11.51\\
\sc spect-test & 5.54 & 57.98 & 57.98 & \textbf{58.01} & 54.76\\
\sc tic-tac-toe & 5.31 & 50.48 & 34.99 & \textbf{56.25} & 6.27\\
\sc zoo-data & 5.85 & 10.66 & 7.37 & \textbf{10.86} & 6.46\\
\midrule
Average & 149.65 & 202.14 & 161.73 & 202.76 & 158.39\\
\bottomrule
\end{tabular}%
\end{table}

\subsection{Results for Uncertainty Measure $R$}

Results for decision trees can be found  in Tables \ref{tab5.1} and \ref{tab5.2}. In
Table \ref{tab5.1}, we consider parameters $h_{R}^{(1)}(T),\ldots ,h_{R}^{(5)}(T)$
(minimum values  are in bold).

\begin{table}[h!]
\caption{Results for $R$ and $h$}
\label{tab5.1}
\begin{tabular}{cccccc}
\toprule  Decision  &   $h_{R}^{(1)}(T)$   &   $h_{R}^{(2)}(T)$ &   $%
h_{R}^{(3)}(T)$ &   $h_{R}^{(4)}(T)$ &   $h_{R}^{(5)}(T)$ \\
 table   $T$  &  &  &  &  &  \\
\midrule
\sc balance-scale   & \bf 4   & \bf 4   & \bf 4   & \bf 4   & \bf 4   \\
\sc breast-cancer   & 6   & 6   & \bf 5   & 6   & \bf 5   \\
\sc cars            & \bf 6   & \bf 6   & \bf 6   & \bf 6   & \bf 6   \\
\sc hayes-roth-data & \bf 4   & \bf 4   & \bf 4   & \bf 4   & \bf 4   \\
\sc lymphography    & \bf 5   & 6   & \bf 5   & 7   & \bf 5   \\
\sc nursery     & 8   & 8   & \bf 7   & 8   & \bf 7   \\
\sc soybean-small   & \bf 2   & 5   & \bf 2   & 7   & \bf 2   \\
\sc spect-test      & 9   & \bf 4   & 5   & 6   & 5   \\
\sc tic-tac-toe     & 7   & \bf 6   & \bf 6   & 8   & 7   \\
\sc zoo-data        & \bf 4   & 5   & \bf 4   & 6   & \bf 4   \\ \midrule
Average         & 5.5 & 5.4 & 4.8 & 6.2 & 4.9 \\
\bottomrule
\end{tabular}%
\end{table}

In Table \ref{tab5.2}, we consider parameters $%
L_{R}^{(1)}(T),\ldots ,L_{R}^{(5)}(T)$ (minimum values  are in bold).

\begin{table}[h!]
\caption{Results for $R$ and $L$}
\label{tab5.2}
\begin{tabular}{cccccc}
\toprule  Decision &   $L_{R}^{(1)}(T)$ &   $L_{R}^{(2)}(T)$ &   $L_{R}^{(3)}(T)$ &   $L_{R}^{(4)}(T)$
&   $L_{R}^{(5)}(T)$ \\
 table   $T$ &  &  &  &  &  \\
\midrule
\sc balance-scale   & \bf 556   & 5,234      & 4,006   & 5234      & 4,006 \\
\sc breast-cancer   & \bf 222   & 19,698     & 320    & 29,308     & 264  \\
\sc cars            & \bf 1,405  & 65,579     & 1,541   & 65,579     & 1,541 \\
\sc hayes-roth-data & \bf 55    & 338       & 65     & 353       & 65   \\
\sc lymphography     & \bf 90    & 47,278     & 110    & 141,322    & 110  \\
\sc nursery      & 4,366  & 12,795,294  & \bf 3,889   & 12,795,294  & \bf 3,889 \\
\sc soybean-small   & \bf 23    & 5,335      & \bf 23     & 36,733     & \bf 23   \\
\sc spect-test      & \bf 53    & 2,730      & 2,126   & 5,920      & 656  \\
\sc tic-tac-toe     & \bf 579   & 164,663    & 74,567  & 645,542    & 734  \\
\sc zoo-data        & \bf 25    & 1,542      & 76     & 3,826      & 72   \\ \midrule
Average         & 737.4 & 1,310,769.1 & 8,672.3 & 1,372,911.1 & 1,136 \\
\bottomrule
\end{tabular}%
\end{table}

Results for decision rules can be found  in Tables \ref{tab5.5} and \ref{tab5.6}. In
Table \ref{tab5.5}, we consider parameters  $l_{R}^{(1)}(T),\ldots ,l_{R}^{(5)}(T)$
(minimum values  are in bold).

\begin{table}[h!]
\caption{Results for $R$ and $l$}
\label{tab5.5}
\begin{tabular}{cccccc}
\toprule  Decision  &   $l_{R}^{(1)}(T)$   &   $l_{R}^{(2)}(T)$ &   $%
l_{R}^{(3)}(T)$ &   $l_{R}^{(4)}(T)$ &   $l_{R}^{(5)}(T)$ \\
 table   $T$  &  &  &  &  &  \\
\midrule
\sc balance-scale & 3.64 & \textbf{3.20} & 3.23 & \textbf{3.20} & 3.23\\
\sc breast-cancer & 3.36 & 2.80 & 3.25 & \textbf{2.79} & 3.33\\
\sc cars & 4.91 & \textbf{2.45} & 4.82 & \textbf{2.45} & 4.82\\
\sc hayes-roth-data & 2.68 & 2.26 & 2.55 &\textbf{ 2.23} & 2.55\\
\sc lymphography & 2.92 & \textbf{2.01} & 3.02 & 2.14 & 3.02\\
\sc nursery & 5.34 & \textbf{3.76} & 5.29 & \textbf{3.76} & 5.29\\
\sc soybean-small & 1.89 & \textbf{1.00} & 1.89 & 1.68 & 1.89\\
\sc spect-test & 3.93 & 2.24 & 1.88 & 2.09 & \textbf{1.84}\\
\sc tic-tac-toe & 5.16 & 3.44 & \textbf{3.43} & 3.44 & 5.16\\
\sc zoo-data & 2.41 & \textbf{1.78} & 2.53 & 2.17 & 2.53\\
\midrule
Average & 3.62 & 2.49 & 3.19 & 2.60 & 3.37\\
\bottomrule
\end{tabular}%
\end{table}

In Table \ref{tab5.6}, we consider parameters $%
c_{R}^{(1)}(T),\ldots ,c_{R}^{(5)}(T)$ (maximum values  are in bold).

\begin{table}[h!]
\caption{Results for $R$ and $c$}
\label{tab5.6}
\begin{tabular}{cccccc}
\toprule  Decision &   $c_{R}^{(1)}(T)$ &   $c_{R}^{(2)}(T)$ &   $c_{R}^{(3)}(T)$ &   $c_{R}^{(4)}(T)$
&   $c_{R}^{(5)}(T)$ \\
 table   $T$ &  &  &  &  &  \\
\midrule
\sc balance-scale & 2.44 & \textbf{4.21} & 4.08 & \textbf{4.21} & 4.08\\
\sc breast-cancer & 4.62 & 7.56 & 4.74 & \textbf{8.14} & 4.81\\
\sc cars & 7.64 & \textbf{332.70} & 8.00 & \textbf{332.70} & 8.00\\
\sc hayes-roth-data & 3.90 & \textbf{6.19} & 3.93 & \textbf{6.19} & 3.93\\
\sc lymphography & 4.53 & 14.91 & 4.72 & \textbf{18.11} & 4.72\\
\sc nursery & 36.52 & \textbf{1,516.81} & 36.77 & \textbf{1,516.81} & 36.77\\
\sc soybean-small & 3.47 & \textbf{12.32} & 3.47 & 10.04 & 3.47\\
\sc spect-test & 20.46 & 55.73 & 55.73 & 56.72 & \textbf{56.75}\\
\sc tic-tac-toe & 12.86 & 25.19 & 25.77 & \textbf{34.12} & 13.10\\
\sc zoo-data & 6.86 & 10.54 & 6.49 & \textbf{10.73} & 6.49\\
\midrule
Average & 10.33 & 198.62 & 15.37 & 199.78 & 14.21\\
\bottomrule
\end{tabular}%
\end{table}

\section{Results of Experiments with Randomly Generated Boolean Functions}

\label{S5}

For $n=3,\ldots ,6$, we randomly generate $100$ Boolean functions with $n$ variables. The table representation $T_f$ of a Boolean function $f(x_1,\ldots ,x_n)$ is considered as a decision table. This table contains $n$ columns and $2^n$ rows. Columns are labeled with variables (attributes) $x_1,\ldots ,x_n$. The set of rows coincides with $\{0,1\}^n$. Each row is labeled with the value of the function $f$ on it.
Decision trees for this decision table are
interpreted as decision trees computing the function $f$.

Using the algorithm $\mathcal{A}_{U}$ with five uncertainty measures, we construct for the generated Boolean functions  decision trees of different types, evaluate complexity of these trees and study decision rules derived from them.

Since each
hypothesis over the decision table $T_f$ is proper, for
each uncertainty measure $U$, $%
h_{U}^{(2)}(T_f)=h_{U}^{(4)}(T_f)$, $h_{U}^{(3)}(T_f)=h_{U}^{(5)}(T_f)$, $L_{U}^{(2)}(T_f)=L_{U}^{(4)}(T_f)
$, $L_{U}^{(3)}(T_f)=L_{U}^{(5)}(T_f)$, $l_{U}^{(2)}(T_f)=l_{U}^{(4)}(T_f)$, $l_{U}^{(3)}(T_f)=l_{U}^{(5)}(T_f)$, $c_{U}^{(2)}(T_f)=c_{U}^{(4)}(T_f)
$, and $c_{U}^{(3)}(T_f)=c_{U}^{(5)}(T_f)$.

The obtained experimental results  for Boolean functions do not depend on the used
uncertainty measures. At the end of this section, we will explain this interesting fact.

Results for decision trees can be found in Tables \ref%
{tab1.3} and \ref{tab1.4}. In Table \ref{tab1.3}, we consider parameters $h_{U}^{(1)},\ldots
,h_{U}^{(5)}$, $U \in \{me,rme,ent,gini,R\}$, in the format $_{\mathit{min}}\mathit{Avg}_{\mathit{max}}$ (minimum values of $\mathit{Avg}$ are in bold).

\begin{table}[h!]
\caption{Results for $U \in \{me,rme,ent,gini,R\}$ and $h$}
\label{tab1.3}
\begin{tabular}{cccccc}
\toprule Number of &   $h_{U}^{(1)}$ &   $h_{U}^{(2)}$ &   $h_{U}^{(3)}$ &   $h_{U}^{(4)}$ &   $%
h_{U}^{(5)} $ \\
variables   $n$&  &  &  &  &  \\
\midrule
    3 & ${}_{0} {2.94}_{3}$ & ${}_{0} {2.02}_{3}$ & ${}_{0} {\bf 1.86}_{3}$ & ${}_{0} {2.02}_{3}$ & ${}_{0} {\bf 1.86}_{3}$  \\
    4 & ${}_{4} {4.00}_{4}$ & ${}_{2} {3.05}_{4}$ & ${}_{2} {\bf 2.97}_{3}$ & ${}_{2} {3.05}_{4}$ & ${}_{2} {\bf 2.97}_{3}$  \\
    5 & ${}_{5} {5.00}_{5}$ & ${}_{4} {4.11}_{5}$ & ${}_{3} {\bf 3.99}_{4}$ & ${}_{4} {4.11}_{5}$ & ${}_{3} {\bf 3.99}_{4}$  \\
    6 & ${}_{6} {6.00}_{6}$ & ${}_{5} {5.09}_{6}$ & ${}_{5} {\bf 5.00}_{5}$ & ${}_{5} {5.09}_{6}$ & ${}_{5} {\bf 5.00}_{5}$ \\
\bottomrule
\end{tabular}%
\end{table}

In Table \ref{tab1.4}, we consider parameters $L_{U}^{(1)},\ldots ,L_{U}^{(5)}$, $U \in \{me,rme,ent,gini,$ $R\}$, in the format $_{%
\mathit{min}}\mathit{Avg}_{\mathit{max}}$ (minimum values of $\mathit{Avg}$ are in bold).

\begin{table}[h!]
\caption{Results for $U \in \{me,rme,ent,gini,R\}$ and $L$}
\label{tab1.4}
\begin{tabular}{cccccc}
\toprule Number of &   $L_{U}^{(1)}$ &   $L_{U}^{(2)}$ &   $L_{U}^{(3)}$ &   $L_{U}^{(4)}$ &   $L_{U}^{(5)} $ \\
variables   $n$ &  &  &  &  &  \\
\midrule
    3 & ${}_{1} {\bf 9.60}_{15}$ & ${}_{1} {12.33}_{22}$ & ${}_{1} {9.61}_{15}$ & ${}_{1} {12.33}_{22}$ & ${}_{1} {9.61}_{15}$  \\
    4 & ${}_{15} {\bf 21.02}_{29}$ & ${}_{14} {44.75}_{70}$ & ${}_{11} {27.69}_{58}$ & ${}_{14} {44.75}_{70}$ & ${}_{11} {27.69}_{58}$  \\
    5 & ${}_{31} {\bf 42.54}_{51}$ & ${}_{125} {218.05}_{292}$ & ${}_{25} {70.19}_{176}$ & ${}_{125} {218.05}_{292}$ & ${}_{25} {70.19}_{176}$  \\
    6 & ${}_{69} {\bf 86.30}_{101}$ & ${}_{649} {1171.03}_{1538}$ & ${}_{75} {292.99}_{807}$ & ${}_{649} {1171.03}_{1538}$ & ${}_{75} {292.99}_{807}$  \\
\bottomrule
\end{tabular}%
\end{table}

Results for decision rules can be found in Tables \ref%
{tab1.7} and \ref{tab1.8}. In Table \ref{tab1.7}, we consider parameters $l_{U}^{(1)},\ldots
,l_{U}^{(5)}$, $U \in \{me,rme,ent,gini,R\}$, in the format $_{\mathit{min}}\mathit{Avg}_{\mathit{max}}$ (minimum values of $\mathit{Avg}$ are in bold).

\begin{table}[h!]
\caption{Results for $U \in \{me,rme,ent,gini,R\}$ and $l$}
\label{tab1.7}
\begin{tabular}{cccccc}
\toprule Number of &   $l_{U}^{(1)}$ &   $l_{U}^{(2)}$ &   $l_{U}^{(3)}$ &   $l_{U}^{(4)}$ &   $%
l_{U}^{(5)} $ \\
variables   $n$ &  &  &  &  &  \\
\midrule
3  & ${}_{1.75}{2.28}_{3.00}$	 &	${}_{1.25}\textbf{{2.09}}_{2.75}$ &	${}_{1.25}{2.10}_{3.00}$  &	${}_{1.25}\textbf{{2.09}}_{2.75}$  &	 ${}_{1.25}{2.10}_{3.00}$ \\

4  &	${}_{2.38}{3.31}_{3.88}$ &	${}_{1.63}\textbf{{2.96}}_{3.69}$ &	${}_{1.63}{3.05}_{3.50}$  &	${}_{1.63}\textbf{{2.96}}_{3.69}$  &	 ${}_{1.63}{3.05}_{3.50}$ \\

5  &	${}_{3.75}{4.29}_{4.63}$ &	${}_{3.25}\textbf{{3.92}}_{4.50}$ &	${}_{3.03}{3.99}_{4.50}$  &	${}_{3.25}\textbf{{3.92}}_{4.50}$  &	 ${}_{3.03}{3.99}_{4.50}$ \\

6 &	${}_{4.94}{5.29}_{5.56}$ &	${}_{4.30}\textbf{{4.88}}_{5.25}$ &	${}_{4.28}{4.89}_{5.34}$  &	${}_{4.30}\textbf{{4.88}}_{5.25}$  &	 ${}_{4.28}{4.89}_{5.34}$ \\
\bottomrule
\end{tabular}%
\end{table}

In Table \ref{tab1.8}, we consider parameters $c_{U}^{(1)},\ldots ,c_{U}^{(5)}$, $U \in \{me,rme,ent,gini,$ $R\}$, in the format $_{%
\mathit{min}}\mathit{Avg}_{\mathit{max}}$ (maximum values of $\mathit{Avg}$ are in bold).

\begin{table}[h!]
\caption{Results for $U \in \{me,rme,ent,gini,R\}$ and $c$}
\label{tab1.8}
\begin{tabular}{cccccc}
\toprule Number of &   $c_{U}^{(1)}$ &   $c_{U}^{(2)}$ &   $c_{U}^{(3)}$ &   $c_{U}^{(4)}$ &   $c_{U}^{(5)} $ \\
variables   $n$ &  &  &  &  &  \\
\midrule 3 & ${}_{1.00}{1.85}_{2.75}$ &	${}_{1.25}\textbf{{2.18}}_{3.63}$ &	${}_{1.00}{2.08}_{3.63}$ &	${}_{1.25}\textbf{{2.18}}_{3.63}$  &	 ${}_{1.00}{2.08}_{3.63}$  \\

4 &	${}_{1.13}{1.85}_{4.63}$ &	${}_{1.31}\textbf{{2.54}}_{6.44}$ &	${}_{1.50}{2.26}_{6.44}$  &	${}_{1.31}\textbf{{2.54}}_{6.44}$   &	 ${}_{1.50}{2.26}_{6.44}$ \\

5 &	${}_{1.38}{1.86}_{3.38}$ &	${}_{1.50}\textbf{{2.69}}_{4.22}$ &	${}_{1.50}{2.36}_{4.84}$  &	${}_{1.50}\textbf{{2.69}}_{4.22}$  &	 ${}_{1.50}{2.36}_{4.84}$ \\

6 &	${}_{1.47}{1.85}_{2.75}$ &	${}_{2.00}\textbf{{2.86}}_{4.17}$ &	${}_{1.67}{2.60}_{4.16}$  &	${}_{2.00}\textbf{{2.86}}_{4.17}$  &	 ${}_{1.67}{2.60}_{4.16}$ \\
\bottomrule
\end{tabular}%
\end{table}

We now explain why the results of experiments with Boolean functions do not
depend on the choice of uncertainty measures from the set $%
M=\{me,rme,ent,$ $gini,R\}$.

Let $f(x_{1},\ldots ,x_{n})$ be a nonconstant Boolean function, $T_{f}$ be
the decision table with attributes $x_{1},\ldots ,x_{n}$ representing $f$,
and $T$ be a nondegenerate subtable of $T_{f}$ such that $%
T=T_{f}\{x_{j_{1}}=\delta _{1},\ldots ,x_{j_{k}}=\delta _{k}\}$, $k<n$, $%
1\leq j_{1}<\cdots <j_{k}\leq n$, and $\delta _{1},\ldots ,\delta _{n}\in
\{0,1\}$.

A subtable $\Theta $ of the table $T_{f}$ is called 1-subtable of $T$ if it
can be represented in the form $\Theta =T\{x_{i}=\delta \}$, where $i\notin
\{j_{1},\ldots ,j_{k}\}$ and $\delta \in \{0,1\}$. This subtable contains
exactly $2^{t}$ rows, where $t=n-k-1$, i.e., $N(\Theta )=2^{t}$. Denote $%
m(\Theta )=\min \{N_{0}(\Theta ),N_{1}(\Theta )\}$. Then

\begin{itemize}
\item $me(\Theta )=m(\Theta )$.

\item $rme(\Theta )=m(\Theta )/2^{t}$.

\item $ent(\Theta )=-p\log _{2}p-(1-p)\log _{2}(1-p)$, where $p=m(\Theta
)/2^{t}$.

\item $gini(\Theta )=m(\Theta )(2^{t}-m(\Theta ))/2^{2t}$.

\item $R(\Theta )=m(\Theta )(2^{t}-m(\Theta ))/2$.
\end{itemize}

We will say that an uncertainty measure $U$ is monotone for $T$ if, for any
1-subtables $\Theta _{1}$ and $\Theta _{2}$ of $T$, $U(\Theta _{1})\leq
U(\Theta _{2})$ if and only if $m(\Theta _{1})\leq m(\Theta _{2})$. We now
show that each uncertainty measure from the set $M$ is monotone for $T$.

For $me$ and $rme$, the considered statement is obvious.

Let us consider the function $H(x)=-x\log _{2}x-(1-x)\log _{2}(1-x)$, where $%
x$ is a real number and $0\leq x\leq 1$. It is well known that this function
is increasing if $0\leq x\leq 0.5$. Using this fact, it is easy to show that
$ent$ is a monotone uncertainty measure for $T$.

We now consider the function $r(x)=x(2^{t}-x)$, where $x$ is an integer and
$0\leq x\leq 2^{t}$. One can show that this function is increasing if $0\leq
x\leq 2^{t-1}$. Using this fact, it is not difficult to show that $gini$ and
$R$ are monotone uncertainty measures for $T$.

Let $U\in M$ and let us assume that, for each pair $\Theta _{1}$, $\Theta
_{2}$ of 1-subtables for $T,$ we know if the inequality $U(\Theta _{1})\leq
U(\Theta _{2})$ holds or not. This information determines the sets of
queries with the minimum impurity among (i) all admissible for $T$
attributes, (ii) all admissible for $T$ hypotheses, and (iii) all admissible
for $T$ attributes and hypotheses. Since all uncertainty measures from the
set $M$ are monotone for $T$, the sets of admissible for $T$ queries with
the minimum impurity do not depend on the choice of uncertainty measures
from $M$.

We will not consider details of the software implementation for the greedy
algorithms. However, the above reasoning allows us to understand
independence of the experimental results for Boolean functions from the
chosen uncertainty measure.

\section{Analysis of Experimental Results}
\label{S6}

First, we evaluate results obtained for the decision tables from \cite{UCI}
based on average values of the parameters $h_{U}^{(k)}$, $L_{U}^{(k)}$, $%
l_{U}^{(k)}$, and $c_{U}^{(k)}$, where $U\in \{me,rme,ent,$ $gini,R\}$ and $%
k=1,\ldots ,5$.

To minimize the depth $h$ of the constructed decision trees, we should
choose decision trees of type 3. The two best uncertainty measures are $R$
and $me$.

To minimize the number of realizable nodes $L$ in the constructed decision
trees, we should choose decision trees of type 1. The two best uncertainty
measures are $R$ and $ent$.

To minimize the length $l$ of decision rules derived from the constructed
decision trees, we should choose decision trees of type 2. The two best
uncertainty measures are $me$ and $rme$.

To maximize the coverage $c$ of decision rules derived from the constructed
decision trees, we should choose decision trees of type 4. The two best
uncertainty measures are $rme$ and $me$.

The results for randomly generated Boolean functions are consistent with the
results for decision tables from \cite{UCI}. For the minimization of $h$, we
should use decision trees of types 3 and 5, for the minimization of $L$ --
trees of type 1, for the minimization of $l$ -- trees of types 2 and 4, and
for the maximization of $c$ -- also trees of types 2 and 4. Note that the
results obtained for Boolean functions do not depend on the choice of an
uncertainty measure from the set $\{me,rme,ent,gini,R\}$.

\section{Conclusions}

\label{S7}

In this paper, we studied decision trees with hypotheses. We designed greedy algorithms based on arbitrary uncertainty
measures for the construction of such decision trees and made experiments
with five uncertainty measures.

Using the results of experiments, we determined which types of decision
trees and uncertainty measures should be chosen if we would like (i) to
minimize the depth of decision trees, (ii) to minimize the number of
realizable nodes in decision trees, (iii) to minimize the length of decision
rules derived from the constructed decision trees, and (iv) to maximize the
coverage of decision rules derived from the constructed decision trees.

\subsection*{Acknowledgments}

Research reported in this publication was supported by King Abdullah
University of Science and Technology (KAUST).

\bibliographystyle{splncs04}
\bibliography{Greedy}
\end{document}